\documentstyle[prl,aps,multicol,float,epsfig]{revtex}
\begin{document}
\draft
\title
{
Two-stage spin-flop transitions in $S$ = 1/2 antiferromagnetic 
spin chain BaCu${}_2$Si${}_2$O${}_7$
}

\author{I. Tsukada,${}^1$ J. Takeya,${}^1$ T. Masuda,${}^{2}$ 
and K. Uchinokura${}^{2}$}
\address{${}^1$Central Research Institute of Electric Power Industry, 
2-11-1 Iwadokita, Komae-shi, Tokyo 201-8511, Japan}
\address{${}^2$Department of Advanced Materials Science, 
The University of Tokyo, 7-3-1 Hongo, Bunkyo-ku, Tokyo 113-8656, Japan}

\date{\today}
\maketitle

\begin{abstract}
Two-stage spin-flop transitions are observed the in quasi-one-dimensional 
antiferromagnet, BaCu${}_2$Si${}_2$O${}_7$. 
A magnetic field applied along the easy axis induces a spin-flop transition 
at 2.0~T followed by a second transition at 4.9~T. 
The magnetic susceptibility indicates the presence of Dzyaloshinskii-Moriya 
(DM) antisymmetric interactions between the intrachain neighboring spins. 
We discuss a possible mechanism whereby the geometrical competition between 
DM and interchain interactions, as discussed for the two-dimensional 
antiferromagnet La${}_2$CuO${}_4$, causes the two-stage spin-flop transitions. 
\end{abstract}
\pacs{75.10.Jm, 75.25.+z, 75.40.Cx, 75.50.Ee}

\begin{multicols}{2}

\narrowtext
The magnetic long-range order (LRO) of a low-dimensional $S$ = 1/2 
antiferromagnet (AF) is qualitatively different from that observed 
in a conventional three-dimensional (3D) AF. 
The absence of LRO in a purely one-dimensional (1D) $S$ = 
1/2 Heisenberg AF at any temperature and a two-dimensional (2D) one at 
$T >$ 0~K was proved several decades ago
\cite{Bethe1}. 
Weak but finite interactions between chains (1D) or planes (2D) are thus 
indispensable for achieving the magnetically ordered state 
at a finite temperature 
\cite{Hutchings1}. 
In such weakly coupled chain or plane systems, the magnitude of 
Dzyaloshinskii-Moriya (DM) antisymmetric interactions, if they exist, 
can be comparable to that of interchain or interplane interactions, 
and as a result, DM interactions cause several behaviors different 
from those of conventional 3D AF. For example, the field-induced gap 
in magnetic excitation has been investigated for 1D Cu benzoate 
\cite{Dender1} 
with respect to the effective staggered field induced by DM interactions 
\cite{Oshikawa1}. 
In this letter, we report exotic two-stage spin-flop transitions 
in the quasi-1D spin system BaCu${}_2$Si${}_2$O${}_7$, 
which are probably due to the competition between 
DM and interchain interactions. 
The magnetic properties of BaCu${}_2M_2$O${}_7$ ($M$ = Si and Ge) have been 
studied recently 
\cite{Tsukada1,Zheludev1,Tsukada2,Kenzelmann1}, 
and it was found that the DM interactions play a central role 
in the weak-ferromagnetic order of BaCu${}_2$Ge${}_2$O${}_7$. 
Since BaCu${}_2$Si${}_2$O${}_7$ is isomorphic to BaCu${}_2$Ge${}_2$O${}_7$, 
we expect the presence of DM interactions in BaCu${}_2$Si${}_2$O${}_7$ also. 
The results are discussed in comparison with the complicated spin-flop 
nature of La${}_2$CuO${}_4$, a parent compound of high-$T_c$ cuprates 
\cite{Thio1}, 
where the symmetric interplane interactions and the antisymmetric intraplane 
DM interactions give rise to a complex spin rotation under magnetic fields.

A single-crystal sample was prepared by a floating-zone (FZ) method as 
previously reported 
\cite{Tsukada1}. 
The crystal was cut into a rectangle with dimensions of 
1$\times$2$\times$1 mm${}^3$ ($a {\times} b {\times} c$). 
The orientations of all the surfaces were confirmed by x-ray Laue 
backscattering. 
Since the diffraction patterns from the $ab$ and $bc$ planes are difficult 
to distinguish, we checked the orientation also using an 
${\omega}$-2${\theta}$ x-ray diffractometer. 
Temperature and magnetic-field dependences of the magnetization 
($M_i(T, H)$, $i$ = $a$, $b$, and $c$) and susceptibility 
($\chi_i(T, H)$ = $\frac{M_i(T, H)}{H})$ were measured using a commercial 
SQUID magnetometer (MPMS-7) up to $H$ = 7~T, 
with which we can control the temperature and magnetic field to within 
$\pm$0.01~K and $\pm$1~Oe of the set values, respectively.

\begin{figure}
\begin{center}
\includegraphics*[width=65mm]{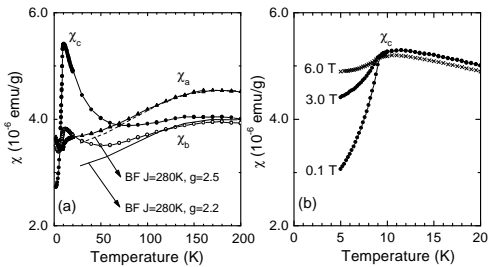}
\end{center}
\caption{
(a) Magnetic susceptibilities along the three principal axes around 
the N{\'e}el temperature measured at $H$ = 0.1~T. The data were taken 
from Ref.~[5]. (b) Low-temperature $c$-axis susceptibilities measured with 
three different fields. $T_N$ is almost unchanged with applied field. 
}
\label{Fig.1}
\end{figure}

First we show the previously reported low-field susceptibility data 
($H$ = 0.1~T) in Fig.~\ref{Fig.1}(a) 
\cite{Tsukada1}. 
The high-temperature behavior obeys the theoretical Bonner-Fisher 
curve 
\cite{Bonner1}, 
indicating paramagnetic behavior of the uniform 1D $S$ = 1/2 Heisenberg 
AF chain. The AF transition occurs at $T_N$ = 9.2~K, below which 
the susceptibilities show anisotropic behavior. 
$\chi_c(T, 0.1~$T$)$ shows a large decrease below $T_N$, while 
$\chi_a(T, 0.1~$T$)$ and $\chi_b(T, 0.1~$T$)$ roughly keep the values 
at $T_N$ down to 2~K. 
This indicates that the $c$ axis is the principal easy axis, 
as confirmed by neutron diffraction 
\cite{Tsukada1,Zheludev1,Kenzelmann1}. 
The large residual susceptibility $\chi_c(2.0~$K$, 0.1~$T$)$ should also 
be noted. The absence of divergent behavior toward $T$ = 0~K 
indicates that this residual $\chi_c$ is not the response of free spins 
due to impurities and/or crystal imperfections, but rather is intrinsic 
to the spin susceptibility of the BaCu${}_2$Si${}_2$O${}_7$ system. 
Moreover, the anisotropy is also observed above $T_N$. 
Two characteristic features are found in this region. 
On the one hand, $\chi_a$ is larger than $\chi_b$ and $\chi_c$ at room 
temperature, which can be attributed to the difference in $g$ values roughly 
determined from the oxygen coordination around Cu${}^{2+}$ ions. 
In the vicinity of $T_N$, on the other hand, the deviation from the 
Bonner-Fisher curve becomes notable only in $\chi_b$ and $\chi_c$, 
while $\chi_a$ retains the Bonner-Fisher-like behavior even close to $T_N$. 
This upward deviation suggests the appearance of an additional contribution 
to the magnetization only along the $bc$ plane.

What characterizes this compound is a peculiar field dependence of 
$\chi_c(T, H)$ below $T_N$. 
As is shown in Fig.~\ref{Fig.1}(b), $\chi_c$ traces three different curves 
at $H$ = 0.1, 3.0, and 6.0~T. 
Since the 0.1~T field is well below the first spin-flop field, as will be 
mentioned later, $\chi_c(T, 0.1~$T$)$ represents the zero-field 
susceptibility. $\chi_c(T, 6.0~$T$)$ shows behavior similar to that of 
conventional AF in the spin-flop phase, where $\chi_c(T)$ maintains 
almost the same value as that at $T_N$. 
In contrast, $\chi_c(T, 3.0~$T$)$ is notable in that it follows 
neither $\chi_c(T, 0.1~$T$)$ nor $\chi_c(T, 6.0~$T$)$. 
It decreases smoothly toward $T$ = 0~K, but not as fast as 
$\chi_c(T, 0.1~$T$)$. 
$T_N$ is almost unchanged and no indication of another transition 
is found. 
Thus we conclude that the three sets of data are intrinsic to 
BaCu${}_2$Si${}_2$O${}_7$.

\begin{figure}
\begin{center}
\includegraphics*[width=55mm]{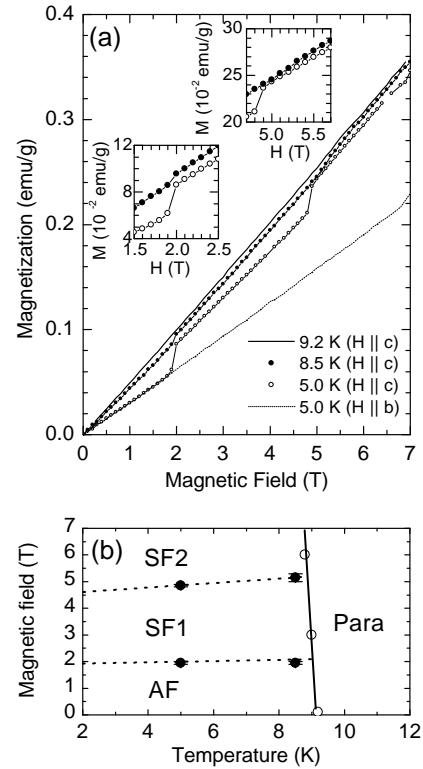}
\end{center}
\caption
{(a) Field dependence of the magnetization along the $c$ axis at $T$ = 5.0, 
8.5, and 9.2~K. At $T$ = 5.0~K, spin-flop transitions are observed at 
$H$ = 2.0 and 4.9~T. The $b$-axis magnetization shows no anomaly. 
(b) Phase diagram around $T_N$. Open and filled circles are the boundaries 
obtained from ${\chi}-T$ and $M-H$ measurements, respectively. 
Solid and dotted lines are guides for the eye.
}
\label{Fig.2}
\end{figure}

The existence of the three phases is more clearly seen in the field 
dependence of $M_c(T, H)$, as shown in Fig.~\ref{Fig.2}(a). 
$M_c(9.2~$K$ (= T_N), H)$ has an almost linear relationship 
with the magnetic field up to $H$ = 7.0~T. 
The data at 8.5~K, which is slightly lower than at $T_N$, shows a steep 
magnetization jump at $H$ = 2.0~T, and broad increase from $H$ = 5.0~T to 
5.3~T. 
Both transitions become clearer in $M_c(5.0~$K$, H)$. 
The magnetization shows a discontinuous jump at $H_{c1}$ = 2.0~T 
and $H_{c2}$ = 4.9~T. 
Consequently, it becomes evident that the $\chi_c$'s shown in 
Fig.~\ref{Fig.1}(b) represent the susceptibilities of these three regions. 
The phase diagram obtained from the magnetization and the susceptibility 
is shown in Fig.~\ref{Fig.2}(b). 
The low-temperature ordered state is separated into three phases; 
for convenience, we call the lower-, middle-, and higher-field phases 
AF, SF1, and SF2, respectively. 
The phase boundaries between AF and SF1 and between SF1 and SF2 
are almost temperature independent. 
We also confirmed that the spin-flop transition is absent when 
the field is applied along the $a$- and $b$-axis directions 
(only $\chi_b(5.0~$K$, H)$ is shown in Fig.~\ref{Fig.2}(a)). 
Thus, we can eliminate the possibility of the mixture of misaligned 
domains being responsible for one of the two transitions.


The presence of three phases is not expected in a conventional AF. 
To construct a proper model in order to explain these two-stage spin-flop 
transitions, we consider the following three results. 
First, the $c$ axis is the principal easy axis at $H$ = 0~T, as was 
determined by neutron diffraction 
\cite{Tsukada1,Kenzelmann1}. 
Second, $\chi_b$ and $\chi_c$ are enhanced in the vicinity of $T_N$. 
Such behavior is characteristic of the AF transition with spin canting, 
as was observed in La${}_2$CuO${}_4$ 
\cite{Thio1,Uchinokura1} 
and Bi${}_2$Sr${}_2$CoO${}_{6.25}$ 
\cite{Tarascon1}. 
In the present case, the canted moments are confined to the $bc$ plane 
and probably originate from the DM interactions. 
Third, $\chi_c(T, 6.0~$T$)$ is almost unchanged below $T_N$. 
Such a weak temperature dependence is usually observed 
in a conventional spin-flop phase, where spins are almost perpendicular 
to the applied field with slight canting toward the field direction. 
In a conventional AF without strong anisotropy, however, the susceptibility 
above $T_N$ exhibits a normal paramagnetic behavior (Curie-Weiss law for 3D and 
Bonner-Fisher curve for 1D), so that the particular enhancement of 
susceptibility is absent. 
Thus the presence of additional susceptibility at 6.0~T below $T_N$ implies 
that spins are more canted toward the $c$ direction than expected for only an 
the external field. The most probable source of this canting is 
DM interactions.


DM interactions often occur in a spin system with low crystallographic 
symmetry. Dzyaloshinskii first pointed out that an asymmetric superexchange 
interactions such as ${\bf D}\cdot{\bf S}_i{\times}{\bf S}_j$ is allowed 
in several antiferromagnets
\cite{Dzyaloshinskii1}. 
Later Moriya gave a microscopic basis for the interactions
\cite{Moriya1}. 
For the particular case of BaCu${}_2$Si${}_2$O${}_7$, the intrachain 
Cu-O-Cu bond actually has a local symmetry (no inversion, but a mirror plane 
including two Cu${}^{2+}$ and one O${}^{2-}$ ion) 
that allows DM interactions at this bond. 
Note that DM interactions are perturbative to the main superexchange 
interaction. Thus it is sufficient to consider DM interactions 
only between the intrachain neighboring Cu${}^{2+}$ ions.

\begin{figure}
\begin{center}
\includegraphics*[width=55mm]{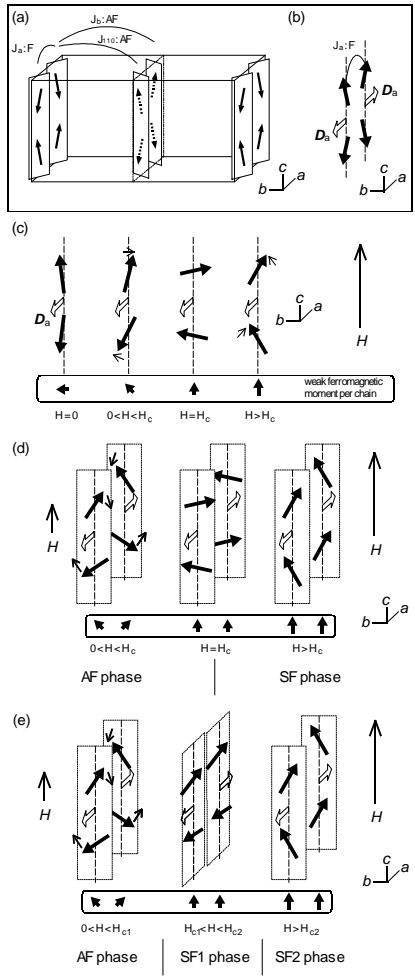}
\end{center}
\caption{
(a) Expected spin arrangement at $H$ = 0~T under the effect of DM + KSEA 
interactions. 
(b) Four spins on the $a$-axis neighboring chains at $H$ = 0~T. 
(c) Spin configurations of a single chain with DM interactions with the 
field applied along the $c$ axis. 
(d) Spin-flop process expected when $J_a$ is negligibly small. 
Spins are confined within the $bc$ plane, and a spin-flop transition occurs 
only at $H_c$. 
(e) Possible two-stage spin-flop transitions. Spins are confined to the $bc$ 
plane in the AF and SF2 phases, while they turn to the $ac$ plane in the SF1 
phase. 
}
\label{Fig.3}
\end{figure}

Let us now construct a model structure, taking DM interactions 
into account. 
By considering the local symmetry of the intrachain Cu-O-Cu bond, 
we can estimate Dzyaloshinskii vectors as real-space vectors 
{\bf D}${}_i  = D((-1)^i0.86, (-1)^i0.51, 0.07)$. 
The magnitude of the $c$ component is one order lower than 
those of the $a$ and $b$ components, so that it can be safely ignored. 
We also assume the presence of an easy-axis anisotropy along the $c$-axis 
as is suggested by $\chi_c(T, 0.1~$T$)$. 
Then the spin arrangement at $H$ = 0~T is obtained as shown in 
Fig.~\ref{Fig.3}(a) under the experimentally determined inter-chain 
interactions: $J$ = 24.1~meV (AF), $J_a$ = -0.460~meV (ferromagnetic), 
$J_b$ = 0.200~meV (AF), and 2$J_{110}$ = 0.152~meV (AF) 
\cite{Kenzelmann1}. 
Spin canting and accompanying weak ferromagnetic moment per chain are present, 
but such weak ferromagnetic moments are mutually compensated between the 
$a$-axis neighboring chains. 
Therefore, this model is consistent with the observation of 
no spontaneous magnetization. 
Whenever we discuss the effect of DM interactions, we should also pay 
attention to the effect of additional pseudo-dipole interactions accompanying 
DM interactions 
\cite{Kaplan1}. 
However, the effect of this term (so-called KSEA interactions) is to erase 
the anisotropy, and under the presence of easy-axis anisotropy 
along the $c$ axis, it does not change the spin configurations shown in 
Fig.~\ref{Fig.3}(a). The effect of this term will also be discussed later.


In a conventional spin-flop theory, only symmetric spin-spin interactions 
are taken into account, and therefore, the competition between 
Zeeman energy and anisotropy energy determines the spin-flop field $H_c$. 
On the other hand, DM interactions are {\em antisymmetric}, 
so that they can compete with symmetric interchain interactions under 
a certain geometry of the spin arrangement. 
This idea was first proposed by Thio {\it et al.} 
to explain the anomalous spin-flop behavior of La${}_2$CuO${}_4$ 
\cite{Thio1}. 
They discussed the magnetic behavior of La${}_2$CuO${}_4$ with respect to 
a staggered moment and a ferromagnetic moment defined as a difference 
and a sum of two neighboring spins belonging to different sublattices. 
When DM interactions are sufficiently strong, the spin-flop transition 
occurs in such a way that the weak-ferromagnetic moments 
(initially perpendicular to the field) rotate toward the field direction, 
and as a result, one does not observe the discontinuous magnetization 
jump at the critical field that is observed in a conventional AF. 
The rotation direction (clockwise or counterclockwise) depends on the 
direction of {\bf D} vectors. 
Thus, if the initial direction of the weak ferromagnetic moments alternate 
with every layer of La${}_2$CuO${}_4$, the rotation directions also 
alternate
\cite{Thio1}. 
Within this rotation process, a competition between DM and interchain 
interactions can occur.

To demonstrate such a competition, 
we further simplify our spin system as follows. 
First, we ignore the effects of $J_b$ and $J_{110}$ and take only $J_a$ 
into account, because $J_a$ is the largest among the interchain interactions. 
Next, we consider the $a$-axis component of {\bf D} vectors ($D_a$) only, 
because $D_a$ must be responsible for the observed $bc$-plane anisotropy. 
Then we can focus our discussion on four particular spins (black arrows) 
shown in Fig.~\ref{Fig.3}(b), 
in which white arrows represent $D_a$. 
Note that these spin chains are coupled with ferromagnetic $J_a$. 
The signs of $D_a$ alternate along the $a$ axis, which means that 
the initial directions of the weak ferromagnetic moments also alternate 
along the $a$-axis. 
When we consider a single chain with alternating $D_a$, as shown in 
Fig.~\ref{Fig.3}(c), magnetic field applied along the $c$ axis causes 
one spin-flop transition at $H_c$. 
Below the spin-flop field, the weak ferromagnetic moment rotates with 
increasing field up to $H_c$, 
which corresponds to each spin also rotating in the same direction.

Now let us take the effect of $a$-axis neighboring chains into account. 
If the effect of $J_a$ is far smaller than that of DM interactions, 
spins may behave in the same way as in the case of a single chain. 
Figure~\ref{Fig.3}(d) shows a schematic of the behavior; 
again we will obtain only one spin-flop field. 
If the magnitude of $J_a$ becomes comparable to that of $D_a$, 
however, the situation differs. 
Because of the alternating $D_a$, the rotation direction of the 
weak ferromagnetic moment (and thus each spin) is opposite to that of the 
$a$-axis neighboring chains. 
In Fig.~\ref{Fig.3}(d), $a$-axis neighboring spins are almost antiparallel 
at $H$ = $H_c$; this is unfavorable for large $J_a$. 
Such an effect seems to be negligible at $H \approx$ 0~T and 
$H \gg$ $H_c$, because in both situations, spins are approximately 
parallel to each other along the $a$-axis. 
Therefore, a new phase can appear only around $H_c$. 
The presence of the SF1 phase suggests that the effect of 
$J_a$ is comparable to that of $D_a$ in BaCu${}_2$Si${}_2$O${}_7$. 
In this phase, the $a$-axis neighboring spins must be parallel to each other. 
There are two possible configurations according 
to the crystal symmetry: 
one is that the spins are pointed approximately in the $a$ direction, 
and the other is that they are pointed approximately in the $b$ direction. 
However, the latter seems to be unfavorable, considering the effect of 
KSEA interactions 
\cite{Kaplan1} 
which effectively add an easy-axis anisotropy along the direction parallel to 
$D_a$. 
This means that spins tend to point in the $a$ direction, and thus 
the former case is favored. 
As a result, we propose a spin configuration of the intermediate SF1 phase, 
as schematically illustrated in Fig.~\ref{Fig.3}(e). 
By increasing magnetic field, spins start to rotate within the $bc$ plane 
(AF) up to $H_{c1}$, switch to the $ac$ plane (SF1), and then return 
to the $bc$ plane (SF2) at $H_{c2}$. 
This concept is consistent with the appearance of additional spin canting 
in the SF2 phase; the alternating $D_a$ can produce additional 
spin canting toward the field direction in every chain. 
Such complex spin-flop transitions cannot be observed in a spin system 
with only symmetric spin-spin interactions, and we believe that they are 
possible only when the symmetric and antisymmetric interactions coexist 
with comparable magnitude.

For further quantitative discussion, we should consider the effects 
of $J_b$, $J_{110}$, and the $b$ component of {\bf D}. 
We intentionally ignore the possibility that spins continuously 
change their directions from the $ac$ to the $bc$ plane. 
This corresponds to assuming the existence of some unknown potential 
barrier between these two configurations. 
However, a recent neutron diffraction study of BaCu${}_2$Si${}_2$O${}_7$ 
in a magnetic field revealed a sudden change of the spin directions, 
which is consistent with the proposed model 
\cite{Ressouche1}. 
The detailed quantitative analysis will be reported in the future.

In summary, we have observed a new type of spin-flop transition, where 
the low dimensionality and accompanying antisymmetric interaction 
cooperatively induce characteristic spin-flop transitions. 
The study of DM interactions is still one of the hot topics of low-dimensional 
quantum antiferromagnets, particularly from a theoretical 
viewpoint \cite{Daniel1}, 
and we believe that this compound will be a well-defined basic material 
with which to study the weakly coupled $S$ = 1/2 spin chain system 
with the DM interaction.

We appreciate A. Zheludev and E. Ressouche for fruitful discussions and for 
sharing their unpublished data of spin structures determined by neutron 
diffraction. We also thank T. A. Kaplan, N. Furukawa and K. Kubo for 
valuable discussions and R. Hiwatari and M. Kiuchi for technical advice. 
Work at the University of Tokyo is supported in part 
by a Grant-in-Aid for COE Research by the Ministry of Education, Culture, 
Sports, Science and Technology of Japan.

\end{multicols}

\end{document}